\documentclass[prd,twocolumn,preprintnumbers,amsmath,amssymb,superscriptaddress,floatfix,nofootinbib]{revtex4}

\usepackage{graphicx}
\usepackage{bm}
\usepackage{booktabs}
\usepackage{amsmath}
\usepackage{amsfonts}
\usepackage{amssymb}
\usepackage{color}
\usepackage{float}
\usepackage{longtable}
\usepackage{tabu}
\usepackage{multirow}
\usepackage[colorlinks, citecolor=blue,anchorcolor=red,menucolor=red, linkcolor=red,filecolor=red,runcolor=red,urlcolor=blue,frenchlinks=red]{hyperref}

\begin{document} 
\title{\Large \bf \boldmath Study of the $B^- \to K^- \eta \eta_c$ decay due to the $D\bar{D}$ bound state} 

\author{Xin-Qiang Li}
\email{xqli@mail.ccnu.edu.cn}
\affiliation{Institute of Particle Physics and Key Laboratory of Quark and Lepton Physics (MOE), Central China Normal University, Wuhan, Hubei 430079, China}
\affiliation{Center for High Energy Physics, Peking University, Beijing 100871, China}

\author{Li-Juan Liu}
\email{liulijuan@zzu.edu.cn}
\affiliation{School of Physics and Laboratory of Zhongyuan Light, Zhengzhou University, Zhengzhou, Henan 450001, China}

\author{En Wang}
\email{wangen@zzu.edu.cn }
\affiliation{School of Physics and Laboratory of Zhongyuan Light, Zhengzhou University, Zhengzhou, Henan 450001, China}
\affiliation{Guangxi Key Laboratory of Nuclear Physics and Nuclear Technology, Guangxi Normal University, Guilin 541004, China}

\author{Le-Le Wei}
\email{llwei@mails.ccnu.edu.cn}
\affiliation{Institute of Particle Physics and Key Laboratory of Quark and Lepton Physics (MOE), Central China Normal University, Wuhan, Hubei 430079, China}

\begin{abstract}
We study the $B^- \to K^- \eta \eta_c$ decay by taking into account the $S$-wave contributions from the pseudoscalar meson--pseudoscalar meson interactions within the unitary coupled-channel approach, where the $D\bar{D}$ bound state is dynamically generated. In addition, the contribution from the intermediate resonance $K_0^*(1430)^-$, with $K_0^*(1430)^- \to K^-\eta$, is also considered. Our results show that there is a clear peak around $3720$~MeV in the $\eta \eta_c$ invariant mass distribution, which could be associated with the $D \bar{D}$ bound state. The future precise measurements of the $B^- \to K^- \eta \eta_c$ process at the Belle II and LHCb experiments could be, therefore, used to check the existence of the $D \bar{D}$ bound state, and to deepen our understanding of the hadron-hadron interactions.
\end{abstract}


\maketitle

\section{Introduction}
\label{Sec:Introduction}

Since the discovery of $X(3872)$ by the Belle Collaboration in 2003~\cite{Belle:2003nnu}, many exotic states, which do not fit into the expectations of conventional quark models, have been observed experimentally during the past two decades~\cite{ParticleDataGroup:2022pth}. Many of these exotic states, especially the ones observed in the charmonium sector, are observed around the threshold of a pair of heavy hadrons; some of them, such as $X(3872)$~\cite{Pakvasa:2003ea}, $Z_{c}(3900)$~\cite{Chen:2015ata} and $X(4160)$~\cite{Molina:2009ct}, can be explained as the hadronic molecules. However, the hadronic molecular states with mass near the $D \bar{D}$ threshold have not yet been observed experimentally, and further detailed studies are therefore required both theoretically and experimentally~\cite{Guo:2017jvc}.
 
In Ref.~\cite{Gamermann:2006nm}, by taking into account the $\pi \pi$, $K \bar{K}$, $D \bar{D}$, $D_s \bar{D}_s$, $\eta \eta$, and $\eta \eta_c$ coupled channels, the authors predicted a narrow hidden charm resonance with quantum numbers $I(J^{PC})=0(0^{++})$ and mass around $3700$~MeV, which will be denoted as $X(3700)$ throughout this paper, within the unitary coupled-channel approach. Furthermore, by considering the $\eta_c$ as a pure $c \bar{c}$ state and the $\eta$--$\eta^\prime$ mixing, together with the same parameters as used in Ref.~\cite{Gamermann:2006nm}, the pole of the new $X(3700)$ state was predicted to be $\sqrt{s}=(3722-i18)~\text{MeV}$ within the unitary coupled-channel approach~\cite{Gamermann:2009ouq}. The mass of the $D \bar{D}$ bound state predicted by other different models is also basically around the $D \bar{D}$ threshold~\cite{Prelovsek:2020eiw,Dong:2021bvy,Chen:2021erj,Shi:2021hzm,Xin:2022bzt,Peng:2023lfw,Mutuk:2022ckn,Wang:2023hpp}, and the theoretical studies of the experimentally measured processes $e^+ e^- \to J/\psi D \bar{D}$~\cite{Belle:2017egg,Gamermann:2007mu,Wang:2019evy}, $B^+ \to D^0 \bar{D}^0 K^+$~\cite{Dai:2015bcc} and $\gamma \gamma \to D \bar {D}$~\cite{Deineka:2021aeu,Belle:2005rte,BaBar:2010jfn,Wang:2020elp} all support the existence of such a $D \bar{D}$ bound state. Meanwhile, some processes like $\psi(3770) \to \gamma X(3700) \to \gamma \eta \eta^\prime$, $\psi(4040) \to \gamma X(3700) \to \gamma \eta \eta^\prime$, $e^+ e^- \to J/\psi X(3700) \to J/\psi \eta \eta^\prime$~\cite{Xiao:2012iq}, $\psi(3770) \to \gamma D \bar{D}$~\cite{Dai:2020yfu}, $\Lambda_b \to \Lambda D \bar{D}$~\cite{Wei:2021usz}, and $B^+ \to K^+ \eta \eta$~\cite{Brandao:2023vyg} have also been suggested to search for the $D \bar{D}$ bound state. It is worth mentioning that the BESIII Collaboration has recently searched for the $X(3700)$ in the $\psi(3770) \to \gamma \eta \eta^\prime$ decay for the first time, observing however no significant signals due to the low detection efficiencies of the photons~\cite{BESIII:2023bgk}.
 
Although the $D\bar{D}$ bound state $X(3700)$ couples mainly to the $D \bar{D}$ and $D_s \bar{D}_s$ channels, it is not easy to search for any signals of the state in these systems. This is due to the fact that, since its mass is a little bit lower than the $D \bar{D}$ threshold, the $X(3700)$ state would manifest itself as a near-threshold enhancement in the $D \bar{D}$ invariant mass distribution, which may be difficult to identify due to the low detection efficiencies near the threshold~~\cite{Wei:2021usz,LHCb:2024hfo}. On the other hand, the $X(3700)$ state has also a sizeable coupling to the $\eta \eta_c$ channel, as observed in Refs.~\cite{Gamermann:2006nm,Gamermann:2009ouq}. Since the $\eta \eta_c$ threshold is about $200$~MeV lower than the predicted mass of $X(3700)$, one expects that, if the $D \bar{D}$ bound state exists, a clear peak near the $D \bar{D}$ threshold would appear in the $\eta \eta_c$ invariant mass distribution of some processes with large phase space.

As is well known, the three-body weak decays of the $B$ mesons involve much more complicated dynamics than do the two-body decays and can, therefore, provide a wealth of information about the meson-meson interactions and the hadron resonances~\cite{Xing:2022uqu,Duan:2023qsg,Lyu:2023jos,Han:2023teq,Huang:2022zsy} (see \textit{e.g.}, Ref.~\cite{Bediaga:2020qxg} for a recent review). For instance, the $B \to K + X/Y/Z$ decay is an ideal process to produce the charmonium-like hadronic molecular states~\cite{Wang:2021aql,Wang:2017mrt,Dai:2018nmw,Zhang:2020rqr,Chen:2021erj}, and many exotic states have been observed experimentally through the $B$-meson weak decays during the past few years, such as $Z_{cs}(4000)$, $Z_{cs}(4220)$~\cite{LHCb:2021uow} and $X(4140)$~\cite{CDF:2009jgo,D0:2013jvp} in $B^+ \to J/\psi \phi K^+$, as well as $X_0(2900)$ and $X_1(2900)$ in $B^+ \to D^+ D^- K^+$ decay~\cite{LHCb:2020bls,LHCb:2020pxc}. In this paper, we propose to search for the $D \bar{D}$ bound state $X(3700)$ in the $B^- \to K^- \eta \eta_c$ decay. It is worth mentioning that the Belle Collaboration has already searched for the process in 2015 based on $772\times10^6$ $B\bar{B}$ pairs collected at the $\Upsilon(4S)$ resonance~\cite{Belle:2015yoa}, but no significant signal of the $D \bar{D}$ bound state was observed due to insufficient statistics. However, the Belle II Collaboration will accumulate about $50$ times the Belle dataset~\cite{Belle-II:2018jsg,Bhardwaj:2018ffc}, and is expected to make further precise measurements of the $B^- \to K^- \eta \eta_c$ decay, which will shed more light on the existence of the $D \bar{D}$ bound state in this process. In addition, the authors of Ref.~\cite{Xie:2022lyw} have suggested to search for the $D \bar{D}$ bound state in the $\eta\eta_c $ mass distribution of the $B^+ \to K^+ \eta \eta_c$ decay, and predicted the branching ratio of $\mathcal{B}(B^+ \to X_{q \bar{q}}(\to \eta_c \eta) K^+)= ( 0.9 \sim 6.7) \times 10^{-4}$.

In this paper, motivated by the observations made above, we will study the $B^- \to K^- \eta \eta_c$ decay by taking into account the pseudoscalar meson--pseudoscalar meson interactions within the chiral unitary approach, from where the $D \bar{D}$ bound state is generated dynamically. On the other hand, the $B^- \to K^- \eta \eta_c$ decay can also proceed through the subsequent decay of the intermediate resonance $K^*_0(1430)$, \textit{i.e.} $K^*_0(1430) \to K \eta$, whose contribution will be  considered in this paper too. We will demonstrate that, besides a peak of $K_0^*(1430)$ in the $K^-\eta$ invariant mass distribution, there is a clear peak around $3720$~MeV in the $\eta \eta_c$ invariant mass distribution, which could be associated with the $D \bar{D}$ bound state. Therefore, future precise measurements of the $B^- \to K^- \eta \eta_c$ decay at the Belle II and LHCb experiments could be used to check the existence of the $D \bar{D}$ bound state, and to deepen our understanding of the hadron-hadron interactions.

This paper is organized as follows. In Sec.~\ref{Sec:Formalism}, we will firstly introduce our formalism for the $B^- \to K^- \eta \eta_c$ decay. Our numerical results and discussions are then presented in Sec.~\ref{Sec:Results}. In Sec.~\ref{Sec:Conclusions}, we give our final conclusion.

\section{Formalism}
\label{Sec:Formalism}

\begin{figure}[htbp]
  \centering
  \includegraphics[scale=0.48]{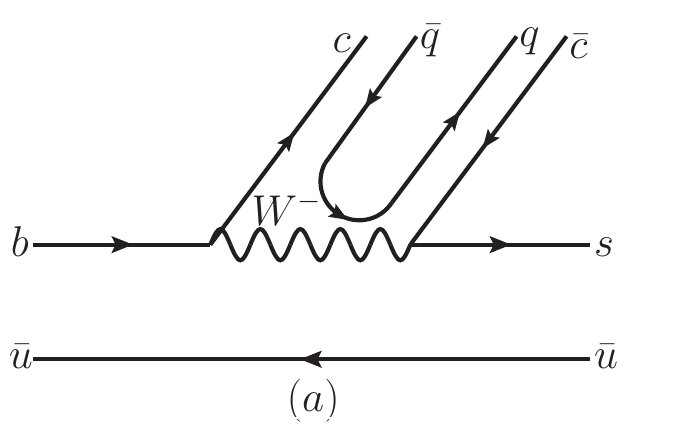}\\
  \vspace{0.15cm}
  \includegraphics[scale=0.48]{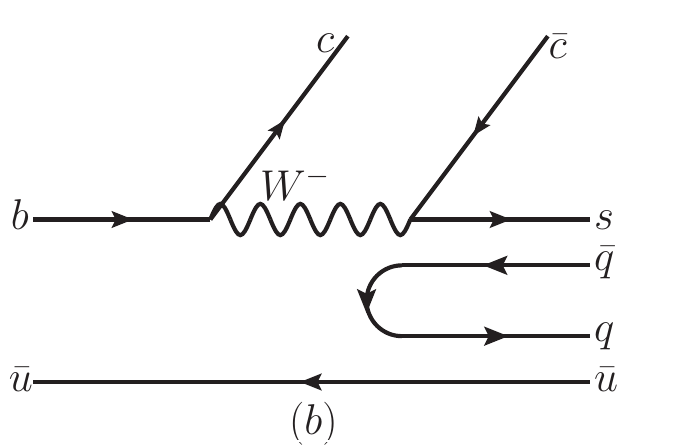}\\
  \vspace{0.15cm}
  \includegraphics[scale=0.48]{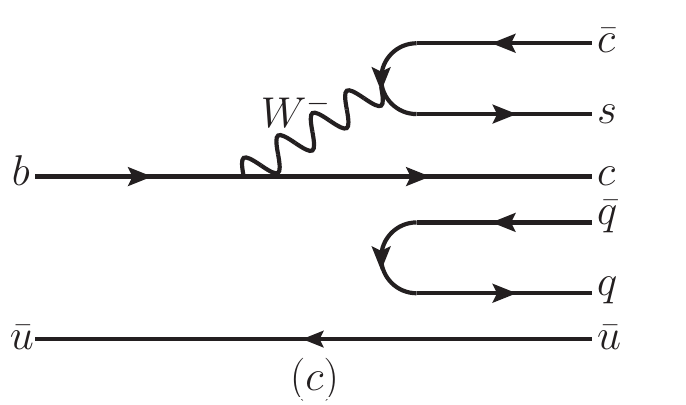}\\
  \vspace{0.15cm}
  \includegraphics[scale=0.48]{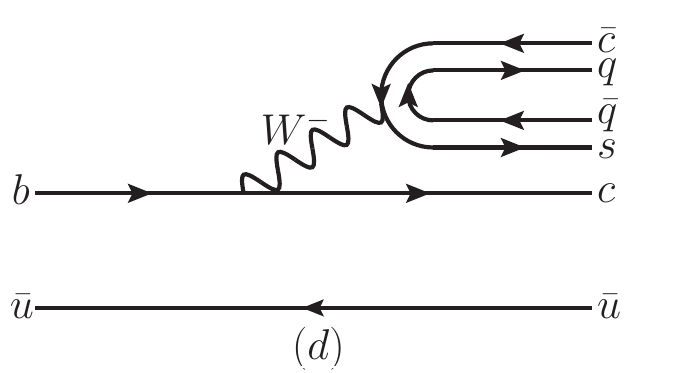}
\caption{The dominant quark-level diagrams for the $B^- \to K^- \eta \eta_c$ decay, where (a)-(b) and (c)-(d) refer to the internal and external $W^-$ emission mechanisms, respectively.}
\label{Fig:QLD}
\end{figure}

In analogy to the discussions made in Refs.~\cite{Wang:2020pem,Wang:2021naf,Liu:2020ajv,Wei:2021usz}, the $B^- \to K^- \eta \eta_c$ decay proceeds via the following three steps: the weak decay, the hadronization, and the final-state interactions. Explicitly, the $b$ quark of the $B^-$ meson firstly decays into a $c$ quark and a virtual $W^{-}$ boson, and then the $W^{-}$ boson turns into a $\bar{c} s$ pair. In order to give rise to the $K^- \eta \eta_c$ final state, the $\bar{u}$ antiquark of the initial $B^-$ meson and the $\bar{c} s$  pair from the $W^-$ subsequent decay have to hadronize together with the $\bar{q} q$ ($\equiv \bar{u} u + \bar{d} d + \bar{s} s$) created from the vacuum with the quantum numbers $J^{PC}=0^{++}$. The relevant quark-level diagrams can be classified as the internal and external $W^-$ emission mechanisms, as depicted in Figs.~\ref{Fig:QLD}(a)--(b) and \ref{Fig:QLD}(c)--(d), respectively. Here we have neglected all the Cabbibo-Kobayashi-Maskawa (CKM) suppressed diagrams that are proportional to the CKM element $V_{ub}$. 

The meson-meson systems formed by the hadronization of $q_i$, $\bar{q}_j$ and $\bar{q}_k q_k$ are given by
\begin{equation}
\sum^3_{k=1}q_i(\bar{q}_k q_k)\bar{q}_j=\sum^3_{k=1}M_{ik}M_{kj}=(M^2)_{ij},
\label{Eq:Hadronization}
\end{equation}
with the $q \bar{q}$ matrix defined as
\begin{equation}
M=\left( \begin{array}{cccc}
           u\bar{u} & u\bar{d} & u\bar{s} & u\bar{c} \\
           d\bar{u} & d\bar{d} & d\bar{s} & d\bar{c} \\
           s\bar{u} & s\bar{d} & s\bar{s} & s\bar{c} \\
           c\bar{u} & c\bar{d} & c\bar{s} & c\bar{c}
         \end{array}
\right),
\label{Eq:qqmatrix}
\end{equation}
which could be expressed in terms of the physical pseudoscalar mesons as~\cite{Lyu:2023jos},
\begin{align}
M &= \nonumber \\
&\hspace{-0.5cm}
 \left( \begin{array}{cccc}
  \frac{\eta}{\sqrt{3}}+\frac{\pi^0}{\sqrt{2}}+\frac{\eta^\prime}{\sqrt{6}} & \pi^+ & K^+ & \bar{D}^0 \\
  \pi^- & \frac{\eta}{\sqrt{3}}-\frac{\pi^0}{\sqrt{2}}+\frac{\eta^\prime}{\sqrt{6}} & K^0 & D^- \\
  K^- & \bar{K}^0 & \sqrt{\frac{2}{3}}\eta^\prime-\frac{\eta}{\sqrt{3}} & D_s^- \\
  D^0 & D^+ & D_s^+ & \eta_c
  \end{array}
\right).\nonumber \\
\label{Eq:SU4matrix}
\end{align}
Thus, by isolating the meson $K^-$, one could easily obtain the components of the meson systems for Figs.~\ref{Fig:QLD}(a) and \ref{Fig:QLD}(b) as follows:
\begin{eqnarray}
\left| H \right\rangle^{\text{a}} &=& V_p V_{cb} V_{cs}^\ast c(\bar{u} u + \bar{d} d + \bar{s} s) \bar{c}s\bar{u}\nonumber\\
&=&V_p V_{cb} V_{cs}^\ast \left(M^2\right)_{44} K^- \nonumber \\
&=& V_p V_{cb} V_{cs}^\ast \times \left( D^0 \bar{D}^0 + D^+ D^- + D_s^+ D_s^- \right) K^-, \nonumber \\ \label{Eq:Ha} \\[0.2cm]
\left| H \right\rangle^{\text{b}}  &=& V_p V_{cb} V_{cs}^\ast c\bar{c}s(\bar{u} u + \bar{d} d + \bar{s} s) \bar{u}\nonumber\\
&=&V_p V_{cb} V_{cs}^\ast \left(M^2\right)_{31} \eta_c \nonumber \\
&=& V_p V_{cb} V_{cs}^\ast \times \left( \frac{1}{\sqrt{2}}K^- \pi^0 + \frac{3}{\sqrt{6}}K^- \eta^\prime  \right) \eta_c, \label{Eq:Hb} 
\end{eqnarray}
where $V_{cb}=0.04182$ and $V_{cs}^*=0.97349$ are the CKM matrix elements, and $V_p$ encodes all the remaining factors arising from the production vertex. Then, the final-state interactions of $D\bar{D}$, $D_s\bar{D}_s$, and $\eta' \eta_c$ will dynamically generate the $D\bar{D}$ bound state, which could decay into the $\eta\eta_c$ system. Here we do not consider the component $K^-\pi^0\eta_c$, since the isospin of the $\pi^0\eta_c$ system is $I=1$.

Similarly, we can write the hadron components for Figs.~\ref{Fig:QLD}(c) and ~\ref{Fig:QLD}(d) that could couple to the $K^-\eta\eta_c$ system as follows:
\begin{eqnarray}
\left| H \right\rangle^{\text{c}} &=& V_p V_{cb} V_{cs}^\ast \times C \times \left( K^- D_s^+ \right) D_s^-, \label{Eq:Hd} \\[0.2cm]
\left| H \right\rangle^{\text{d}} &=& V_p V_{cb} V_{cs}^\ast \times C \times \left( K^- \bar{D}^0 \right) D^0, \label{Eq:He}
\end{eqnarray}
where we have introduced the color factor $C$ to account for the relative weight of the external $W^-$ emission mechanism with respect to the internal $W^-$ emission mechanism, and will take $C=3$ in the case of color number $N_C=3$, as done in Refs.~\cite{Duan:2020vye,Zhang:2022xpf,Feng:2020jvp}.

\begin{figure}[htbp]
\includegraphics[width=6.5cm]{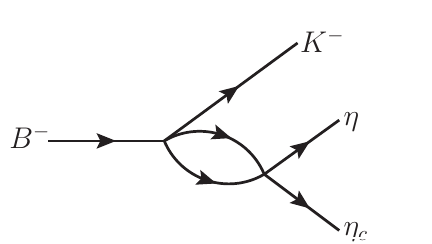}
\caption{The final-state interactions of the coupled channels $D^0 \bar{D}^0$, $D^+ D^-$, $D_s^+ D_s^-$, and $\eta'\eta_c$.}
\label{Fig:Loop}
\end{figure}

According to the above discussions, the $K^- \eta \eta_c$ final state could not be produced directly through the tree-level diagrams of the $B^-$ decay, but can via the final-state interactions of the coupled channels $D^0 \bar{D}^0$, $D^+ D^-$, $D_s^+ D_s^-$, and $\eta'\eta_c$, which could then generate the $D\bar{D}$ bound state, as shown in Fig.~\ref{Fig:Loop}. The total amplitude of Fig.~\ref{Fig:Loop} can be expressed as
\begin{eqnarray}
\mathcal{T}_{X} &=& V_p V_{cb} V_{cs}^\ast \biggl[ G_{D^+ D^-}\, t_{D^+ D^- \to \eta \eta_c}  \nonumber \\
&& \qquad\qquad \left. + (1+C) \times G_{D^0 \bar{D}^0}\, t_{D^0 \bar{D}^0 \to \eta \eta_c} \right. \nonumber \\[0.20cm]
&& \qquad\qquad \left. + (1+C) \times G_{D_s^+ D_s^-}\, t_{D_s^+ D_s^- \to \eta \eta_c} \right. \nonumber \\
&& \qquad\qquad + \frac{3}{\sqrt{6}}\times G_{\eta'\eta_c}\, t_{\eta'\eta_c \to \eta \eta_c} \biggr],
\label{Eq:T1}
\end{eqnarray}
where $G_l$ is the loop function for the two-meson propagation in the $l$-th channel, and its explicit expression is given by~\cite{Gamermann:2006nm}
\begin{eqnarray}
G_{l} &=& i \int \frac{d^4 q}{(2\pi)^4} \frac{1}{q^2 - m_1^2 + i\epsilon} \frac{1}{(P-q)^2 - m_2^2 + i\epsilon} \nonumber \\
&=& \frac{1}{16\pi^2} \left[\alpha_l + \ln{\frac{m_1^2}{\mu^2}} + \frac{m_2^2 - m_1^2 + s}{2s} \ln{\frac{m_2^2}{m_1^2}} \right. \nonumber \\
&& \qquad + \frac{p}{\sqrt{s}} \times \left(\ln{\frac{s - m_2^2 + m_1^2 + 2p\sqrt{s}}{-s + m_2^2 - m_1^2 + 2p \sqrt{s}}} \right. \nonumber \\
&& \qquad \qquad \left. \left. + \ln{\frac{s + m_2^2 - m_1^2 + 2p\sqrt{s}}{-s - m_2^2 + m_1^2 + 2p \sqrt{s}}} \right) \right],
\label{Eq:LoopFuntion}
\end{eqnarray}
with the subtraction constant $\alpha_l= -1.3$ for the coupled channels $D^{+} D^{-}$, $D^0 \bar{D}^0$, $D_s^{+} D_s^{-}$, and $\eta^\prime \eta_c$, and $\mu=1500$~MeV, being the same as used in Ref.~\cite{Gamermann:2009ouq}. $\sqrt{s}=M_{\eta \eta_c}$ is the invariant mass of the two mesons in the $l$-th channel, and $m_1$ and $m_2$ are the masses of these two mesons. $P$ is the total four-momentum of the two mesons in the $l$-th channel, and $p$ is the magnitude of the three-momentum of each meson in the meson-meson center of mass frame, with
\begin{equation} \label{Eq:ThreeMomentum}
p = \frac{\lambda^{1/2} \left( s, m_1^2, m_2^2 \right)}{2 \sqrt{s}},
\end{equation}
where $\lambda(x,y,z) = x^2 + y^2 + z^2 - 2xy - 2yz -2zx$ is the K\"{a}llen function. The transition amplitudes in Eq.~(\ref{Eq:T1}) are obtained by solving the Bethe-Salpeter equation in coupled channels~\cite{Gamermann:2006nm,Gamermann:2009ouq},
\begin{equation} \label{Eq:BSE}
t = [1-VG]^{-1} V,
\end{equation}
where the matrix $V$ is the potential constructed at the tree level for each one of the possible channels. Here we take into account the channels of $\pi^+ \pi^-$, $\pi^0 \pi^0$, $K^+ K^-$, $K^0 \bar{K}^0$, $\eta \eta$, $\eta \eta_c$, $D^+ D^-$, $D^0 \bar{D}^0$, $D_s^+ D_s^-$, $\eta \eta^\prime$, $\eta^\prime \eta^\prime$, as well as $\eta^\prime \eta_c$, and present the transition potential $V_{ij}$ in Table~\ref{Table:Potentials} of App.~\ref{app:potential}.

\begin{figure}[htbp]
\includegraphics[width=8.6cm]{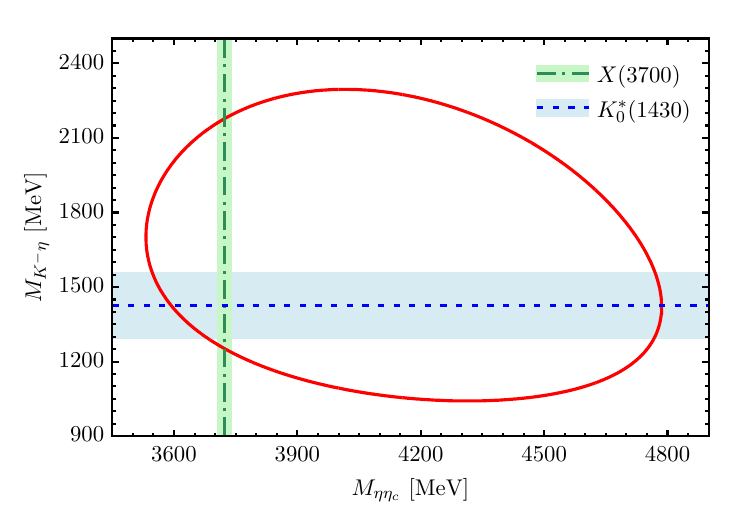}
\caption{The Dalitz plot for the $B^- \to K^- \eta \eta_c$ decay. The green dash-dotted line and band stand for the mass and width of $X(3700)$, while the blue dashed line and band for the mass and width of the well-established resonance $K^*_0(1430)$.}
\label{Fig:BmDalitz}
\end{figure}

\begin{figure}[htbp]
\includegraphics[width=6.8cm]{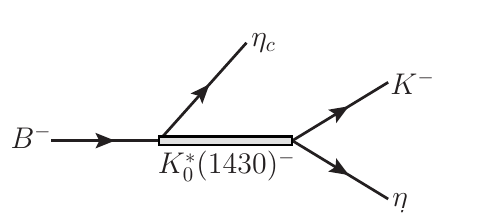}
\caption{The diagram for the $B^- \to K^- \eta \eta_c$ decay via the intermediate $K_{0}^{*}(1430)$ resonance.}
\label{Fig:K1430}
\end{figure}

On the other hand, the $B^- \to K^- \eta \eta_c$ decay could also proceed via the intermediate excited kaon mesons. According to the Dalitz plot shown in Fig.~\ref{Fig:BmDalitz}, one can see that only the well-established resonance $K^*_0(1430)$ could contribute to this process, since the $K^*_0(1430)$ couples to the channel $K^-\eta$ in $S$-wave with a branching fraction $\mathcal{B}(K^*_0(1430)\to K\eta)=(8.6^{+2.7}_{-3.4})\%$~\cite{ParticleDataGroup:2022pth}. Therefore, in this paper, we  neglect all the other excited kaon mesons, and only take into account the contribution from the intermediate $K^*_0(1430)$ resonance as shown by Fig.~\ref{Fig:K1430}, whose amplitude can be expressed as
\begin{equation}
\mathcal{T}_{K^*_0} = \frac{V_p \times \beta \times e^{i\varphi} \times M_{K^*_0(1430)}^2}{M_{K^- \eta}^2 - M_{K^*_0(1430)}^2 + i M_{K^*_0(1430)} \Gamma_{K^*_0(1430)}},
\label{Eq:T2}
\end{equation}
where the parameter $\beta$ accounts for the relative weight of the $K^*_0(1430)$ contribution with respect to that of the $D\bar{D}$ bound state $X(3700)$, and the phase factor $e^{i\varphi}$ is introduced to describe the interference between the amplitudes from the $D\bar{D}$ bound state and the $K_{0}^{*}(1430)$ resonance. $M_{K^- \eta}$ is the invariant mass of the $K^- \eta$ system. We will take as input $M_{K^*_0(1430)} = 1425~\text{MeV}$ and $\Gamma_{K^*_0(1430)} = 270~\text{MeV}$~\cite{ParticleDataGroup:2022pth}.

With the amplitudes given by Eqs.~(\ref{Eq:T1}) and (\ref{Eq:T2}) at hand, the doubly differential decay width of the $B^- \to K^- \eta \eta_c$ process can be written as
\begin{eqnarray}
\frac{\mathrm{d}^2 \Gamma}{\mathrm{d}M_{\eta \eta_c} \mathrm{d}M_{K^- \eta}} &=& \frac{1}{(2 \pi)^3} \frac{M_{\eta \eta_c} M_{K^- \eta}}{8 M_{B^-}^3}{\left|\mathcal{T}_{X} + \mathcal{T}_{K^*_0} \right|}^2, \label{eq:dw1}\\
\frac{\mathrm{d}^2 \Gamma}{\mathrm{d}M_{\eta \eta_c} \mathrm{d}M_{K^- \eta_c}} &= &\frac{1}{(2 \pi)^3} \frac{M_{\eta \eta_c} M_{K^- \eta_c}}{8 M_{B^-}^3}{\left|\mathcal{T}_{X} + \mathcal{T}_{K^*_0} \right|}^2.\label{eq:dw2}
\end{eqnarray}
One could obtain the invariant mass distributions $d\Gamma/dM_{\eta \eta_c}$, $d\Gamma/dM_{K^-\eta}$, and $d\Gamma/dM_{K^-\eta_c}$ by integrating Eqs.~(\ref{eq:dw1}) and (\ref{eq:dw2}) over each of the invariant mass variables.  For instance, the differential decay width ${\mathrm{d}\Gamma}/{\mathrm{d}M_{\eta \eta_c}}$ can then be obtained by integrating Eq.~(\ref{eq:dw1}) over the $K^- \eta$ invariant mass $M_{K^- \eta}$, with the final result given by
\begin{align} \label{Eq:Gamma1}
\frac{\mathrm{d} \Gamma}{ \mathrm{d} M_{\eta \eta_c} } = \int \mathrm{d}M_{K^- \eta} \frac{1}{(2 \pi)^3} \frac{M_{\eta \eta_c} M_{K^- \eta}}{8 M_{B^-}^3}{\left|\mathcal{T}_{X} + \mathcal{T}_{K^*_0} \right|}^2. 
\end{align}
Here the integration range is given by
\begin{align}
& \left( M^2_{K^- \eta} \right)_{\mathrm{min}} \nonumber \\
&= \left( E_{K^-}^* + E_{\eta}^* \right)^2 - \left( \sqrt{E_{\eta}^{*2} - m_{\eta}^2} + \sqrt{E_{K^-}^{*2} - m_{K^-}^2} \right)^2, \label{Eq:Limit1} \\
& \left( M^2_{K^- \eta} \right)_{\mathrm{max}} \nonumber \\
&= \left( E_{K^-}^* + E_{\eta}^* \right)^2 - \left( \sqrt{E_{\eta}^{*2} - m_{\eta}^2} - \sqrt{E_{K^-}^{*2} - m_{K^-}^2} \right)^2, \label{Eq:Limit2}
\end{align}
where $E_{K^-}^*$ and $E_{\eta}^*$ are the energies of $K^-$ and $\eta$ in the $\eta \eta_c$ rest frame, respectively. Explicitly, we have
\begin{eqnarray}
E_{K^-}^* &=& \frac{M^2_{B^-} - M^2_{\eta \eta_c} - M^2_{K^-}}{2 M_{\eta \eta_c}}, \label{Eq:E1} \\[0.15cm]
E_{\eta}^* &=& \frac{M^2_{\eta \eta_c} - M^2_{\eta_c} + M^2_{\eta}}{2 M_{\eta \eta_c}}. \label{Eq:E2}
\end{eqnarray}
Here all the meson masses involved are taken from Ref.~\cite{ParticleDataGroup:2022pth}.

\section{Results and Discussion}
\label{Sec:Results}

\begin{figure*}[ht]
  \centering
  \includegraphics[width=8.6cm]{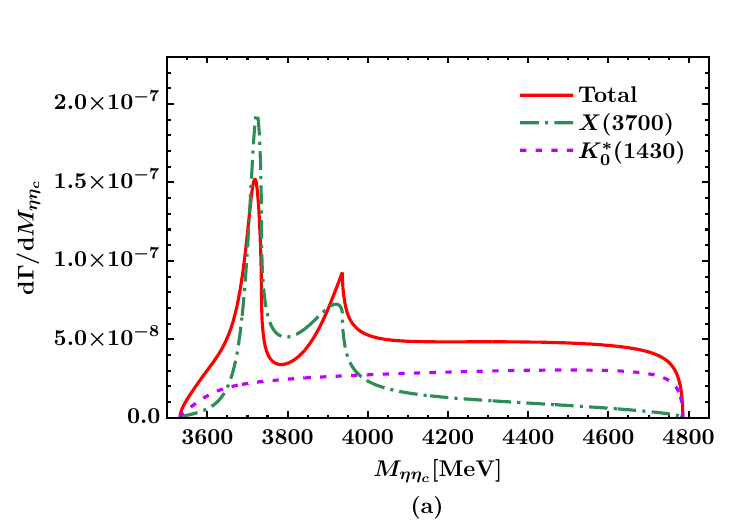}
  \includegraphics[width=8.6cm]{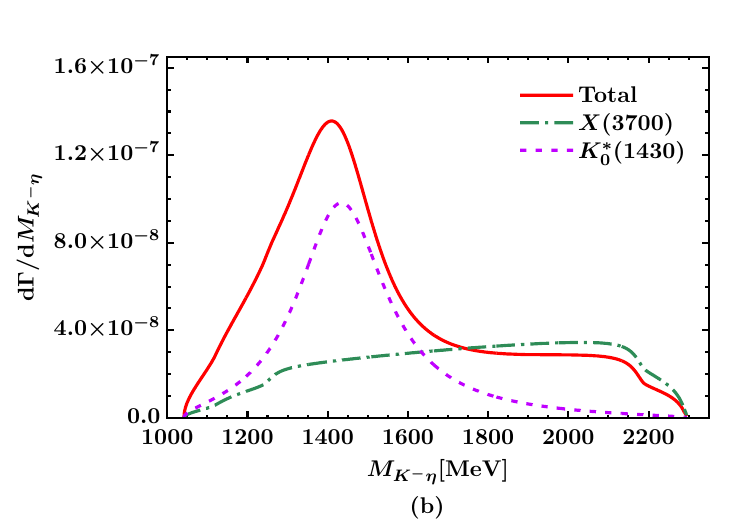}
  \includegraphics[width=8.6cm]{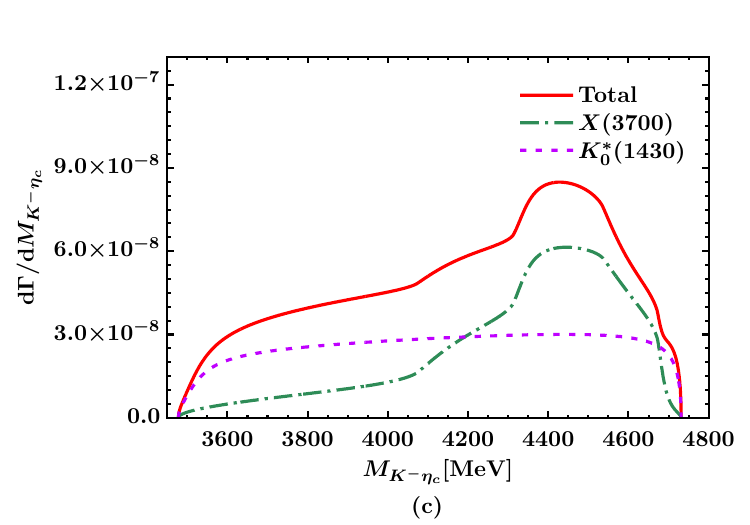}
  \caption{The $\eta \eta_c$ (a), $K^- \eta$ (b) and $K^- \eta_c$ (c) invariant mass distributions of the $B^- \to K^- \eta \eta_c$ decay with $\beta=0.012$, $\varphi=0$ and $C=3.0$. The green dash-dotted, the magenta dashed, and the red solid curves represent the contributions from $X(3700)$, $K^*_0(1430)$, and the total contributions, respectively.}
  \label{Fig:IMD}
\end{figure*}

\begin{figure}[ht]
  \centering
  \includegraphics[width=0.5\textwidth]{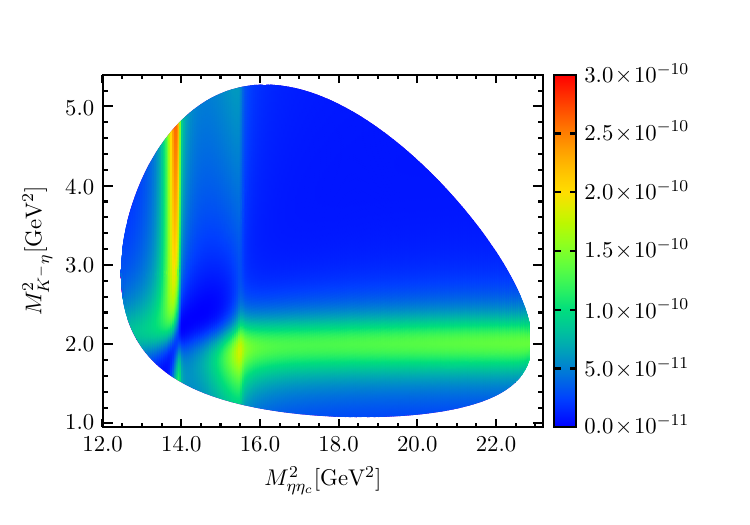}
  \caption{The doubly differential decay width ${\mathrm{d}^2 \Gamma }/\left(\mathrm{d} M_{\eta \eta_c} \mathrm{d} M_{K^- \eta}\right)$ of the $B^- \to K^- \eta \eta_c$ decay in the $(M_{\eta \eta_c}^2, M_{K^- \eta}^2)$ plane, where the $X(3700)$ and $K^*_0(1430)$ resonances can be clearly seen.}
  \label{Fig:Dalitz}
\end{figure}

\begin{figure*}[htbp]
  \centering
  \includegraphics[width=8.6cm]{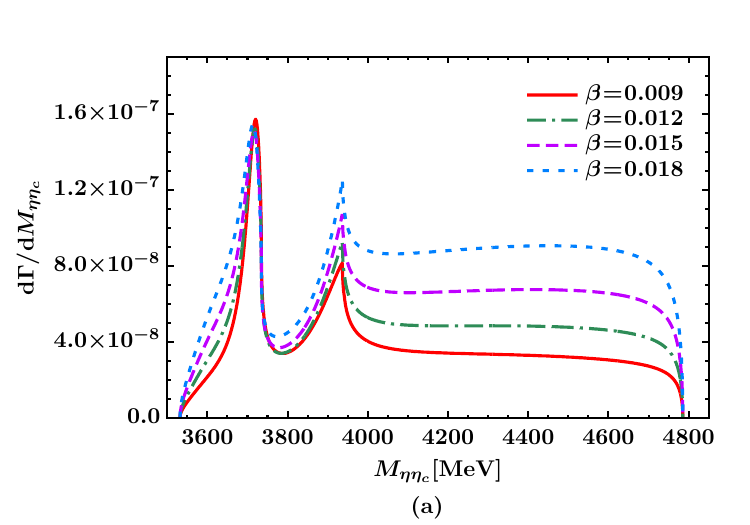}
  \includegraphics[width=8.6cm]{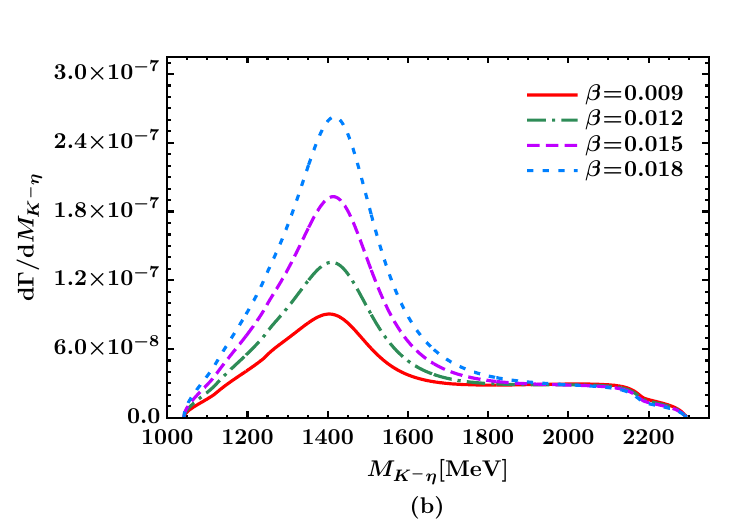}
  \includegraphics[width=8.6cm]{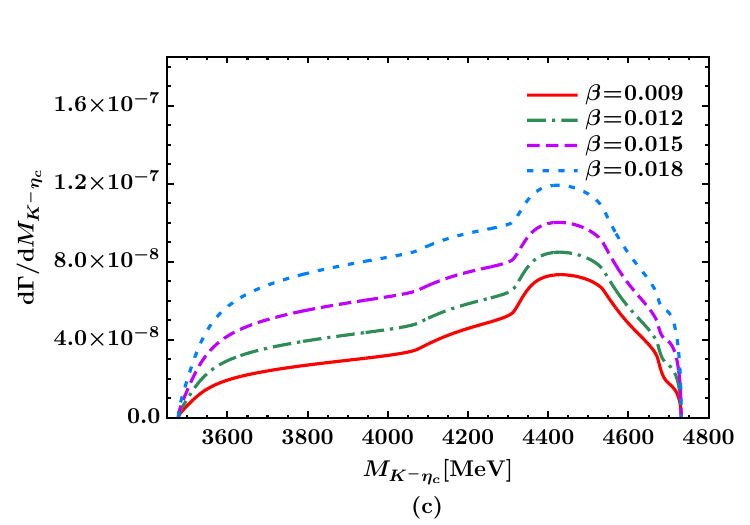}
  \caption{The $\eta \eta_c$ (a), $K^- \eta$ (b) and $K^- \eta_c$ (c) invariant mass distributions of the $B^- \to K^- \eta \eta_c$ decay with $\varphi=0$, $C=3.0$, as well as four different values of $\beta = 0.009$ (red solid), $0.012$ (green dash-dotted), $0.015$ (magenta long-dashed) and $0.018$ (blue dashed).}
  \label{Fig:Beta}
\end{figure*}

\begin{figure*}[htbp]
  \centering
  \includegraphics[width=8.6cm]{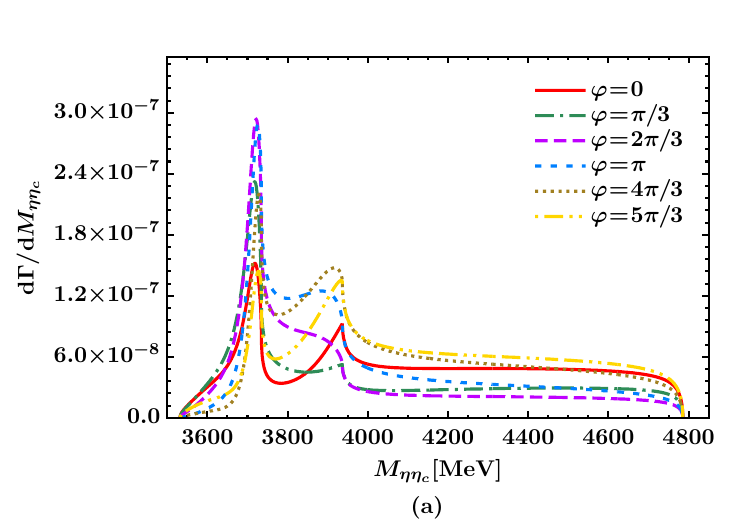}
  \includegraphics[width=8.6cm]{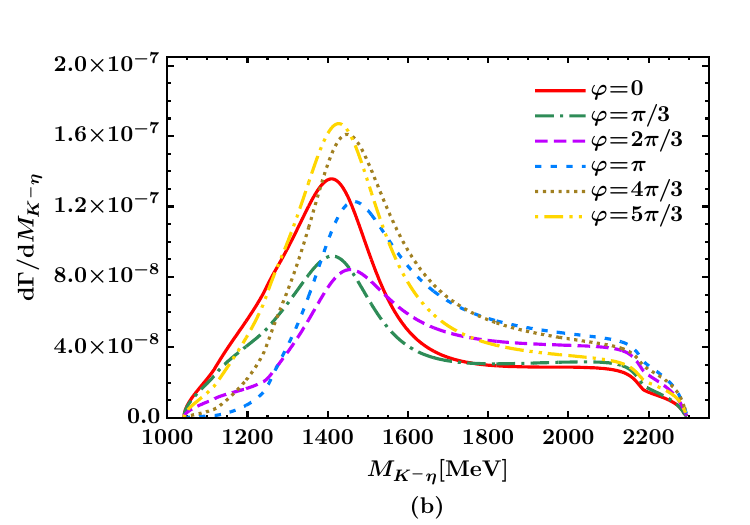}
  \includegraphics[width=8.6cm]{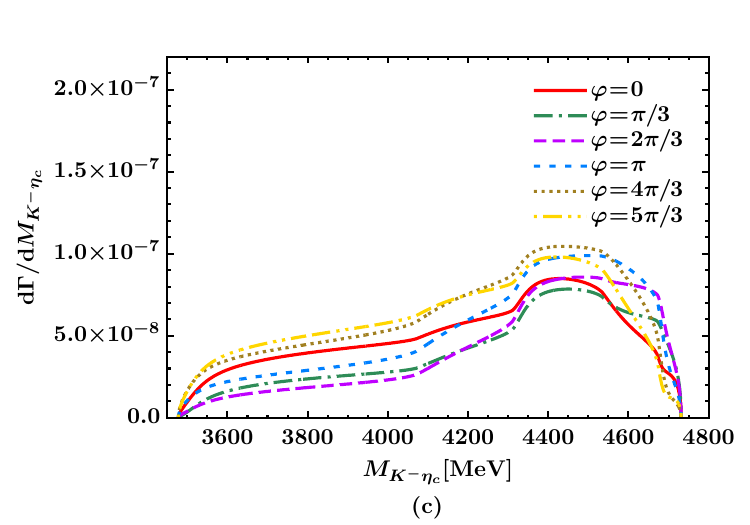}
  \caption{The $\eta \eta_c$ (a), $K^- \eta$ (b) and $K^- \eta_c$ (c) invariant mass distributions of the $B^- \to K^- \eta \eta_c$ decay with $\beta=0.012$, $C=3.0$, as well as six different values of $\varphi = 0$ (red solid), $\pi/3$ (green dash-dotted), $2\pi/3$ (magenta long-dashed), $\pi$ (blue dashed), $4\pi/3$ (olive dotted) and $5\pi/3$ (yellow dash-dot-dotted).}
  \label{Fig:Phase}
\end{figure*}

\begin{figure*}[htbp]
  \centering
  \includegraphics[width=8.6cm]{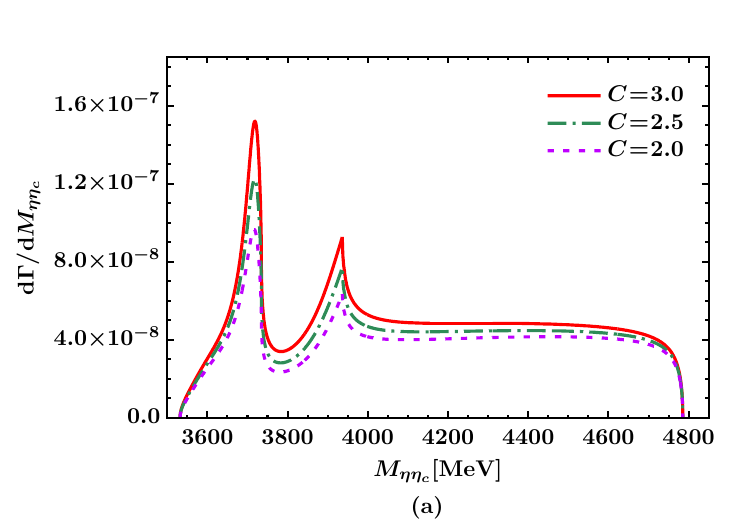}
  \includegraphics[width=8.6cm]{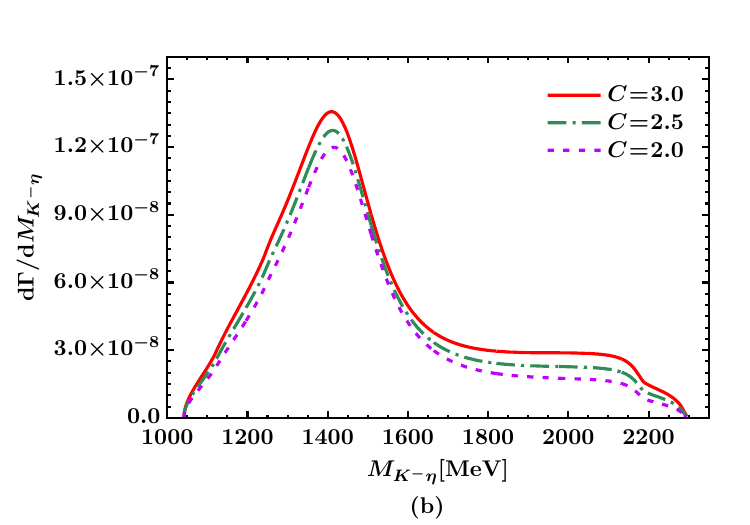}
  \includegraphics[width=8.6cm]{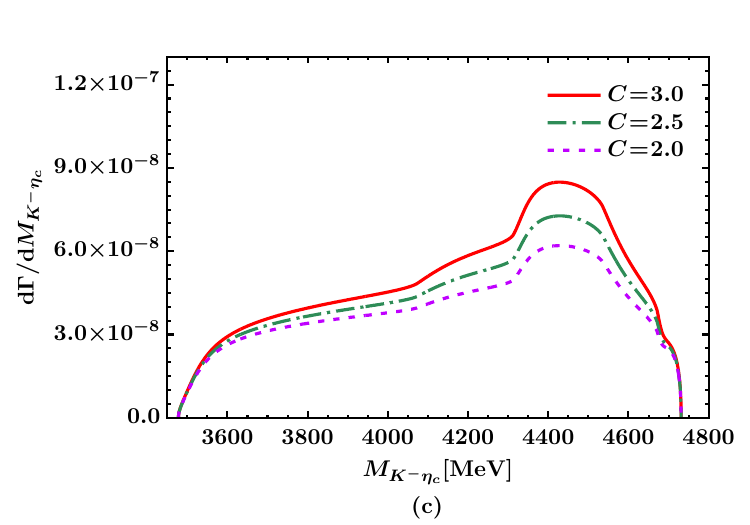}
  \caption{The $\eta \eta_c$ (a), $K^- \eta$ (b) and $K^- \eta_c$ (c) invariant mass distributions of the $B^- \to K^- \eta \eta_c$ decay with $\beta=0.012$, $\varphi=0$, as well as three different values of $C = 3.0$ (red solid), $2.5$ (green dash-dotted), and $2.0$ (magenta dashed).}
  \label{Fig:C}
\end{figure*}

In our model, we have three free parameters, $V_p$, $\beta$ and $\varphi$. The parameter $V_p$ is a global factor and its value does not affect the shapes of the $\eta \eta_c$, $K^- \eta$, \textcolor{red}{and $K^-\eta_c$} invariant mass distributions, and thus we take $V_p=1$ for simplicity. The parameter $\beta$ represents the relative weight of the $K^*_0(1430)$ contribution with respect to that of $X(3700)$, and the parameter $\varphi$ is the relative phase between these two amplitudes.

As indicated by the current data on the branching fractions of $B$-meson decays~\cite{ParticleDataGroup:2022pth}, 
\begin{eqnarray}
\mathcal{B}(B^0 \to K^*_0(1430)^0 \eta_c) &=& \left( 1.8 \pm 0.4 \right) \times 10^{-4}, \label{Eq:Br1} \nonumber\\
\mathcal{B}(B^0 \to K^0 D^+ D^-) &=& \left( 7.5 \pm 1.7 \right) \times 10^{-4}, \label{Eq:Br2} \nonumber\\
\mathcal{B}(B^0 \to K^0 D^0 \bar{D}^0) &=& \left( 2.7 \pm 1.1 \right) \times 10^{-4}, \nonumber \label{Eq:Br3} \\
\mathcal{B}(B^+ \to K^+ D^+ D^-) &=& \left( 2.2 \pm 0.7 \right) \times 10^{-4}, \label{Eq:Br4} \nonumber\\
\mathcal{B}(B^+ \to K^+ D^0 \bar{D}^0) &=& \left( 1.45 \pm 0.33 \right) \times 10^{-3},\nonumber\label{Eq:Br5}
\end{eqnarray}
the branching fractions of the processes $B^0 \to K^*_0(1430)^0 \eta_c$ and $B^0 \to K^0 D \bar{D}$ are of the same order of magnitude. Thus, the contributions from the $D\bar{D}$ bound state and the $K_0^*(1430)$ resonance are expected to be of similar magnitudes. By integrating the differential decay width over the corresponding invariant mass, one can estimate the partial decay widths $\Gamma(B^- \to K^*_0(1430)^- \eta_c \to K^- \eta \eta_c)$ and $\Gamma(B^- \to K^- X(3700) \to K^- \eta \eta_c)$. It is found numerically that, with $\beta=0.012$, the values of $\Gamma(B^- \to K^*_0(1430)^- \eta_c \to K^- \eta \eta_c)$ and $\Gamma({B^- \to K^- X(3700) \to K^- \eta \eta_c})$ are of the same order of magnitude. Therefore, in this work, we take the parameter $\beta=0.012$ and also discuss our results with different values of $\beta$ later.

Firstly, we show in Fig.~\ref{Fig:IMD} the $\eta \eta_c$, $K^- \eta$, and $K^-\eta_c$ invariant mass distributions with $\beta=0.012$ and $\varphi=0$. One can see a clear peak around $3720$~MeV in the $\eta \eta_c$ invariant mass distribution, which should be associated with the $D \bar{D}$ bound state $X(3700)$. At the same time, a cusp structure appears around $3930$~MeV in the same invariant mass distribution, which is due to the strong coupling of the $D\bar{D}$ bound state to the $D_s\bar{D}_s$ channel. In addition, a $K^*_0(1430)$ signal appears in the $K^- \eta$ invariant mass distribution, but gives rise to a smooth shape in the $\eta \eta_c$ invariant mass distribution and thus does not affect the peak structure of the $X(3700)$ significantly. It should be stressed that the line shape of the $X(3700)$ in the $\eta\eta_c$ invariant mass distribution is different from that of a Breit-Wigner form, which is a typical feature of the $D\bar{D}$ molecular state. On the other hand, one bump structure appears around $4400$~MeV in the $K^-\eta_c$ invariant mass distribution, which is due to the $D\bar{D}$ interaction and hence should not be associated with any resonance.

It is worth mentioning that one narrow state $\chi_{c0}(3930)$, with mass around $3930$~MeV and the quantum numbers $J^{PC}=0^{++}$, was observed in the process $B^+ \to D^+ D^- K^+$ process by the LHCb Collaboration~\cite{LHCb:2020pxc}. In addition, the LHCb Collaboration has discovered a peaking structure near the $D_s^+ D_s^-$ threshold, denoted as $X(3960)$ with $M=3956 \pm 5 \pm 10$~MeV, $\Gamma=43 \pm 13 \pm 8$~MeV and $J^{PC}=0^{++}$, in the process $B^+ \to D_s^+ D_s^- K^+$~\cite{LHCb:2022aki}. Some studies suggest that the near-threshold structure, $X(3960)$, may come from the $D_s^+ D_s^-$ bound state below the $D_s^+ D_s^-$ threshold, which can be associated with the $\chi_{c0}(3930)$ state~\cite{Guo:2022zbc,Ding:2023yuo,Liu:2021xje,Bayar:2022dqa,Xin:2022bzt,Lu:2023mgx,Qi:2023gwb,Li:2023wxm,Chen:2023eix}. Taking into account the fact that the $\chi_{c0}(3930)$ and $X(3960)$ states favor the same quantum number $J^{PC}=0^{++}$ and both of them can decay into the $\eta \eta_c$ final state, the cusp structure around $3930$~MeV in the $\eta\eta_c$ invariant mass distribution could be associated with the resonance $X(3930)$. Therefore, the future precise measurements of this process could be used to search for the $\chi_{c0}(3930)$ and $X(3960)$ states. 

We also show in Fig.~\ref{Fig:Dalitz} the doubly differential decay width ${\mathrm{d}^2 \Gamma}/\left(\mathrm{d} M_{\eta \eta_c} \mathrm{d} M_{K^- \eta}\right)$ for the $B^- \to K^- \eta \eta_c$ decay in the $(M_{\eta \eta_c}^2, M_{K^- \eta}^2)$ plane, where one can see two clear bands corresponding to the $X(3700)$ and $K^*_0(1430)$ resonances, respectively.

The parameters $\beta$ and $\varphi$ are unknown in our model, and their values could be determined if the precise experimental measurements of the $B^- \to K^- \eta \eta_c$ decay are available in the future. In order to study the dependence of our results on $\beta$ and $\varphi$, we have calculated the $\eta \eta_c$, $K^- \eta$, and $K^-\eta_c$ invariant mass distributions with different values of $\beta$ and $\varphi$, which are shown in Figs.~\ref{Fig:Beta} and \ref{Fig:Phase}, respectively. From Fig.~\ref{Fig:Beta}, one can see that the peak of the $K^*_0(1430)$ resonance in the $K^- \eta$ invariant mass distribution becomes more significant when the value of $\beta$ increases. From Fig.~\ref{Fig:Phase}, on the other hand, the peak of $K_0^*(1430)$ moves a little bit for different values of $\varphi$. However, the peak of the $D \bar{D}$ bound state $X(3700)$ is always clear in the $\eta \eta_c$ invariant mass distribution.
   
Finally, we should note that the value of the color factor $C$, which represents the relative weight of the external $W^-$ emission mechanism with respect to the internal $W^-$ emission mechanism, could vary around $3$ in order to account for the potential nonfactorizable contributions~\cite{Ali:1998eb}. To this end, we show in Fig.~\ref{Fig:C} the $\eta \eta_c$, $K^- \eta$, and $K^-\eta_c$ invariant mass distributions of the $B^- \to K^- \eta \eta_c$ decay by taking three different values of $C = 3.0$, $2.5$, $2.0$. One can see that, although the peak of the $X(3700)$ state in the $\eta\eta_c$ invariant mass distribution becomes weaker when the value of $C$ decreases, its signal is still clear and can be easily distinguished from the background contribution. Meanwhile, the peak of the $K^*_0(1430)$ resonance in the $K^-\eta$ invariant mass distribution has little changes for these three different values of $C$, because the contribution from the $D\bar{D}$ bound state is smooth around the peak of $K^*_0(1430)$ in the $K^-\eta$ invariant mass distribution, as observed already in Fig.~\ref{Fig:IMD}. 

From the above analyses, one can conclude that, within the variation ranges of the three free parameters, there is always a clear peak around $3720$~MeV in the $\eta \eta_c$ invariant mass distribution, which corresponds to the $D \bar{D}$ bound state. Thus, we strongly suggest our experimental colleagues to perform more precise measurements of the $B^- \to K^- \eta \eta_c$ decay at the Belle II and LHCb experiments in the future, which is very important for confirming the existence of the predicted $D \bar{D}$ bound state.

\section{Conclusions}
\label{Sec:Conclusions}

In this paper, motivated by the theoretical predictions for the $D\bar{D}$ bound state $X(3700)$, we propose to search for this state in the $B^- \to K^- \eta \eta_c$ decay. To this end, we have investigated the process within the unitary coupled-channel approach, by taking into account the contributions from the $S$-wave pseudoscalar meson--pseudoscalar meson interactions, which can dynamically generate the $D\bar{D}$ bound state $X(3700)$. We have also taken into account the contribution from the intermediate resonance $K^*_0(1430)$, since it couples to the $K\eta$ channel in $S$ wave with a branching fraction of $\mathcal{B}(K^*_0(1430)\to K\eta)=(8.6^{+2.7}_{-3.4})\%$.

Our results show that a clear peak appears around $3720$~MeV in the $\eta \eta_c$ invariant mass distribution, which should be associated with the $D\bar{D}$ bound state. It should be stressed that the line shape of the $D\bar{D}$ bound state is significantly different from that of a Breit-Winger form, which is a typical feature of the $D\bar{D}$ molecular state. On the other hand, one can also find the peak of the resonance $K^*_0(1430)$ in the $K^-\eta$ invariant mass distribution, and the resonance gives a smooth contribution in the $\eta\eta_c$ invariant mass distribution.

In summary, we strongly encourage our experimental colleagues to perform a more precise measurement of the $B^- \to K^- \eta \eta_c$ decay at the Belle II and LHCb experiments in the future, which will be very helpful to confirm the existence of the predicted $D \bar{D}$ bound state, as well as to deepen our understanding of the hadron-hadron interactions.

\section*{Acknowledgements}

This work is supported by the Natural Science Foundation of Henan under Grant No. 222300420554 and No. 232300421140, the National Natural Science Foundation of China under Grant No. 12135006, No, 12075097, and No. 12192263, the Project of Youth Backbone Teachers of Colleges and Universities of Henan Province (2020GGJS017),  the Open Project of Guangxi Key Laboratory of Nuclear Physics and Nuclear Technology (No. NLK2021-08), as well as the Fundamental Research Funds for the Central Universities under Grant Nos. CCNU19TD012 and CCNU22LJ004.

{\appendix 
\section{Detailed derivations of the potential $V$}
\label{app:potential}

Following Refs.~\cite{Gamermann:2006nm,Gamermann:2009ouq}, we can write the interaction Lagrangian as
\begin{equation}
\mathcal{L}_{\text{int}}=\frac{1}{12f^2} \text{Tr} \left[ J_{\mu} J^{\mu} + \mathcal{M} \Phi^4\right],
\label{Eq:Lagrangian1}
\end{equation}
where the current $J^{\mu}$ is defined by
\begin{eqnarray}
J^{\mu}=\left( \partial^{\mu} \Phi \right) \Phi - \Phi \left( \partial^{\mu} \Phi \right),
\end{eqnarray}
and the matrix $\mathcal{M}$ of the mass term reads 
\begin{eqnarray} 
\mathcal{M}=\left( \begin{array}{cccc}
           m_{\pi}^2 & 0 & 0 & 0 \\
           0 & m_{\pi}^2 & 0 & 0 \\
           0 & 0 & 2m_K^2-m_{\pi}^2 & 0 \\
           0 & 0 & 0 & 2m_D^2-m_{\pi}^2
         \end{array}
         \right).
\label{Eq:MassMatrix}
\end{eqnarray}
In the physical basis, the matrix $\Phi$ can be written as~\cite{Gamermann:2009ouq}
\begin{widetext}
\begin{eqnarray}
\Phi = \left( \begin{array}{cccc}
 \frac{\pi^0}{\sqrt{2}}+ \frac{\eta}{\sqrt{3}}+\frac{\eta^\prime}{\sqrt{6}} & \pi^+ & K^+ & \bar{D}^0 \\
  \pi^- & \frac{\eta}{\sqrt{3}}-\frac{\pi^0}{\sqrt{2}}+\frac{\eta^\prime}{\sqrt{6}} & K^0 & D^- \\
  K^- & \bar{K}^0 & \sqrt{\frac{2}{3}}\eta^\prime-\frac{\eta}{\sqrt{3}} & D_s^- \\
  D^0 & D^+ & D_s^+ & \eta_c
  \end{array}
\right).
\label{Eq:SU4matrix}
\end{eqnarray}
\end{widetext}
which can be further decomposed into
\begin{eqnarray}
&&\phi_8 = \nonumber \\
&&\left( \begin{array}{ccc}
 \frac{\pi^0}{\sqrt{2}}+ \frac{\eta}{\sqrt{3}}+\frac{\eta^\prime}{\sqrt{6}} & \pi^+ & K^+ \\
  \pi^- & -\frac{\pi^0}{\sqrt{2}}+\frac{\eta}{\sqrt{3}}+\frac{\eta^\prime}{\sqrt{6}} & K^0 \\
  K^- & \bar{K}^0 & -\frac{\eta}{\sqrt{3}} +\sqrt{\frac{2}{3}}\eta^\prime
  \end{array}
\right), \nonumber \\
\label{Eq:phi8}
\end{eqnarray}
\begin{eqnarray}
\phi_{\bar{3}} = \left(\begin{array}{ccc}
D^0 & D^+ & D_s^+ 
\end{array}
\right),
\label{Eq:phi3bar}
\end{eqnarray}
\begin{eqnarray}
\phi_3 = \left( \begin{array}{ccc}
\bar{D}^0 \\
D^- \\
D_s^- 
\end{array}
\right),\label{Eq:phi3}
\end{eqnarray}
\begin{eqnarray}
\phi_1 = \eta_c.
\label{Eq:phi1}
\end{eqnarray}

Since the $\text{SU(4)}$ flavor symmetry breaking already arises from the mass term $\mathcal{M}$, which is not proportional to the identity matrix~\cite{Gamermann:2009ouq}, we are going to use the meson decay constant $f = f_{\pi} = 93~\text{MeV}$ for light mesons, and $f = f_D = 165~\text{MeV}$ for heavy ones. In addition, we are going to suppress all the terms in the Lagrangian where the interaction is driven by the exchange of a heavy vector meson, as usually discussed in the vector meson dominance picture. For details of this suppression, one could refer to Refs.~\cite{Gamermann:2006nm,Gamermann:2007fi,Hofmann:2005sw,Mizutani:2006vq}. Finally, the full corrected Lagrangian can be written as,
\begin{widetext}
\begin{eqnarray}
\mathcal{L} &=& \frac{1}{12 f^2}\left\{ \operatorname{Tr}\left[ J_{88 \mu} J_{88}^{\mu}+2 J_{3 \overline{3} \mu} J_{88}^{\mu}+J_{3 \overline{3}_{\mu}} J_{3 \overline{3}}^{\mu} \right] +\frac{8}{3} \gamma J_{\overline{3} 1 \mu} J_{13}^{\mu} \right. \nonumber \\
&& \qquad \quad +\frac{4}{\sqrt{3}} \gamma\left(J_{\overline{3} 1 \mu} J_{83}^{\mu}+J_{\overline{3} 8 \mu} J_{13}^{\mu} \right) \left. +2 \gamma J_{\overline{3} 8 \mu} J_{83}^{\mu}+\psi_{5} J_{\overline{3} 3 \mu} J_{\overline{3} 3}^{\mu}+\mathcal{L}_{\text {mass}} \right\},
\label{Eq:Lagrangian2}
\end{eqnarray}
\end{widetext}
with
\begin{eqnarray}
\mathcal{L}_{\text {mass}}=\operatorname{Tr}\left[ \mathcal{M} \Phi^4 \right],
\end{eqnarray}
and the correction parameters given by
\begin{eqnarray}
\gamma &=& \left( \frac{m_L}{m_H} \right)^2, \\
\psi_3 &=& \frac{1}{3} + \frac{2}{3} \left( \frac{m_L}{m_{J/\psi} } \right)^2, \\
\psi_5 &=& -\frac{1}{3} + \frac{4}{3} \left( \frac{m_L}{m_{J/\psi}} \right)^2.
\label{Eq:psi}
\end{eqnarray}
Here, $m_L$ and $m_H$ are the masses of light and heavy vector mesons respectively, and they will be set to $m_L=800~\rm{MeV}$ and $m_H=2050~\rm{MeV}$~\cite{Gamermann:2006nm}. When inserting these amplitudes in the Bethe-Salpeter equation, we should divide the amplitude by ${1}/{\sqrt{2}}$ each time when the initial or the final state contains a pair of identical particles (unitary normalization) in order to ensure closure of the intermediate states.

In this paper, we take into account the coupled channels $\pi^+ \pi^-$, $\pi^0 \pi^0$, $K^+ K^-$, $K^0 \bar{K}^0$, $\eta \eta$, $\eta \eta_c$, $D^+ D^-$, $D^0 \bar{D}^0$, $D_s^+ D_s^-$, $\eta \eta^\prime$, $\eta^\prime \eta^\prime$, and $\eta' \eta_c$, and present the transition potential $V_{ij}$ in Table~\ref{Table:Potentials}.

{
\renewcommand\arraystretch{1.5}
\begin{longtable*}{cc}
\caption{The transition potentials $V_{ij}$ among different channels, where $s$, $t$ and $u$ are the Mandelstam variables.}
\label{Table:Potentials} \\
\hline \hline
Channel & Potential \\
\hline
\endfirsthead
\multicolumn{2}{l}
{- continued from previous page} \\
\hline 
Channel & Potential \\
\hline
\endhead
\hline
\multicolumn{2}{l}{continued on next page} \\
\endfoot
\hline \hline
\endlastfoot
$\pi^+ \pi^- \to \pi^+ \pi^-$ & $-\frac{1}{3f^2} \left( (s+t-2u) + 2m_{\pi}^2 \right)$ \\
$\pi^+ \pi^- \to \pi^0 \pi^0$ & $-\frac{1}{3f^2} \left( (2s-t-u) + m_{\pi}^2 \right)$  \\
$\pi^+ \pi^- \to K^+ K^-$ & $-\frac{1}{6f^2} \left( (s+t-2u) + m_{K}^2 + m_{\pi}^2 \right)$ \\
$\pi^+ \pi^- \to K^0 \bar{K}^0$ & $-\frac{1}{6f^2} \left( (s+u-2t) + m_{K}^2 + m_{\pi}^2 \right)$ \\
$\pi^+ \pi^- \to \eta \eta$ & $-\frac{2}{3f^2} ~ m_{\pi}^2$ \\
$\pi^+ \pi^- \to \eta \eta_c$ & $0$ \\
$\pi^+ \pi^- \to D^+ D^-$ & $-\frac{1}{6f^2} \left( (t-u) + \gamma(s-u) + m_D^2 + m_{\pi}^2 \right)$ \\
$\pi^+ \pi^- \to D^0 \bar{D}^0$ & $-\frac{1}{6f^2} \left( (u-t) + \gamma(s-t) + m_D^2 + m_{\pi}^2 \right)$ \\
$\pi^+ \pi^- \to D_s^+ D_s^-$ & $0$ \\
$\pi^+ \pi^- \to \eta \eta^\prime$ & $-\frac{\sqrt{2}}{3f^2} ~ m_{\pi}^2$ \\
$\pi^+ \pi^- \to \eta^\prime \eta^\prime$ & $-\frac{1}{3f^2} ~ m_{\pi}^2$ \\
$\pi^+ \pi^- \to \eta^\prime \eta_c$ & $0$ \\
 & \\
$\pi^0 \pi^0 \to \pi^0 \pi^0$ & $-\frac{1}{f^2} ~ m_{\pi}^2$ \\
$\pi^0 \pi^0 \to K^+ K^-$ & $-\frac{1}{12f^2} \left( (2s-t-u) + 2m_K^2 + 2m_{\pi}^2 \right)$ \\
$\pi^0 \pi^0 \to K^0 \bar{K}^0$ & $-\frac{1}{12f^2} \left( (2s-t-u) + 2m_K^2 + 2m_{\pi}^2 \right)$ \\
$\pi^0 \pi^0 \to \eta \eta$ & $-\frac{2}{3f^2} ~ m_{\pi}^2$ \\
$\pi^0 \pi^0 \to \eta \eta_c$ & $0$ \\
$\pi^0 \pi^0 \to D^+ D^-$ & $-\frac{1}{12f^2} \left( \gamma(2s-t-u) + 2m_D^2 + 2m_{\pi}^2\right)$ \\
$\pi^0 \pi^0 \to D^0 \bar{D}^0$ & $-\frac{1}{12f^2} \left( \gamma(2s-t-u) + 2m_D^2 + 2m_{\pi}^2\right)$ \\
$\pi^0 \pi^0 \to D_s^+ D_s^-$ & $0$ \\
$\pi^0 \pi^0 \to \eta \eta^\prime$ & $-\frac{\sqrt{2}}{3f^2} ~ m_{\pi}^2$ \\
$\pi^0 \pi^0 \to \eta^\prime \eta^\prime$ & $-\frac{1}{3f^2} ~ m_{\pi}^2$ \\
$\pi^0 \pi^0 \to \eta^\prime \eta_c$ & $0$ \\
 & \\
$K^+ K^- \to K^+ K^-$ & $-\frac{1}{3f^2} \left( (s+t-2u) + 2m_K^2 \right)$ \\
$K^+ K^- \to K^0 \bar{K}^0$ & $-\frac{1}{6f^2} \left( (s+t-2u) + 2m_K^2 \right)$ \\
$K^+ K^- \to \eta \eta$ & $-\frac{2}{9f^2} \left( (2s-t-u) + m_K^2\right)$ \\
$K^+ K^- \to \eta \eta_c$ & $0$ \\
$K^+ K^- \to D^+ D^-$ & $0$\\
$K^+ K^- \to D^0 \bar{D}^0$ & $-\frac{1}{6f^2} \left( (u-t) + \gamma(s-t) + m_D^2 + m_K^2 \right)$ \\
$K^+ K^- \to D_s^+ D_s^-$ & $-\frac{1}{6f^2} \left( (t-u) + \gamma(s-u) + m_D^2 + 2m_K^2 - m_{\pi}^2 \right)$ \\
$K^+ K^- \to \eta \eta^\prime$ & $\frac{\sqrt{2}}{18f^2} \left( (2s-t-u) + 4m_K^2 - 3m_{\pi}^2 \right)$ \\
$K^+ K^- \to \eta^\prime \eta^\prime$ & $-\frac{1}{36f^2} \left( (2s-t-u) + 34m_K^2 - 6m_{\pi}^2 \right)$ \\
$K^+ K^- \to \eta^\prime \eta_c$ & $0$ \\
 & \\
$K^0 \bar{K}^0 \to K^0 \bar{K}^0$ & $-\frac{1}{3f^2} \left( (s+t-2u) + 2m_K^2 \right)$ \\
$K^0 \bar{K}^0 \to \eta \eta$ & $-\frac{2}{9f^2} \left( (2s-t-u) +  m_K^2\right)$ \\
$K^0 \bar{K}^0 \to \eta \eta_c$ & $0$ \\
$K^0 \bar{K}^0 \to D^+ D^-$ & $-\frac{1}{6f^2} \left( (u-t) + \gamma(s-t) + m_D^2 + m_K^2 \right)$ \\
$K^0 \bar{K}^0 \to D^0 \bar{D}^0$ & $0$ \\
$K^0 \bar{K}^0 \to D_s^+ D_s^-$ & $-\frac{1}{6f^2} \left( (t-u) + \gamma(s-u) + m_D^2 + 2m_K^2 - m_{\pi}^2 \right)$ \\
$K^0 \bar{K}^0 \to \eta \eta^\prime$ & $\frac{\sqrt{2}}{18f^2} \left( (2s-t-u) + 4m_K^2 - 3m_{\pi}^2 \right)$ \\
$K^0 \bar{K}^0 \to \eta^\prime \eta^\prime$ & $-\frac{1}{36f^2} \left( (2s-t-u) + 34m_K^2 - 6m_{\pi}^2 \right)$ \\
$K^0 \bar{K}^0 \to \eta^\prime \eta_c$ & $0$ \\
 & \\
$\eta \eta \to \eta \eta$ & $-\frac{2}{9f^2} \left( 2m_K^2 + m_{\pi}^2 \right)$ \\
$\eta \eta \to \eta \eta_c$ & $0$ \\
$\eta \eta \to D^+ D^-$ & $-\frac{1}{18f^2} \left( \gamma(2s-t-u) + 2m_D^2 + 2m_{\pi}^2\right)$ \\
$\eta \eta \to D^0 \bar{D}^0$ & $-\frac{1}{18f^2} \left( \gamma(2s-t-u) + 2m_D^2 + 2m_{\pi}^2\right)$ \\
$\eta \eta \to D_s^+ D_s^-$ & $-\frac{1}{18f^2} \left( \gamma(2s-t-u) + 2m_D^2 + 6m_K^2 - 4m_{\pi}^2\right)$ \\
$\eta \eta \to \eta \eta^\prime$ & $\frac{4\sqrt{2}}{9f^2} \left( m_K^2 - m_{\pi}^2 \right)$ \\
$\eta \eta \to \eta^\prime \eta^\prime$ & $-\frac{2}{9f^2} \left( 4m_K^2 - m_{\pi}^2 \right)$ \\
$\eta \eta \to \eta^\prime \eta_c$ & $0$ \\
 & \\
$\eta \eta_c \to \eta \eta_c$ & $0$ \\
$\eta \eta_c \to D^+ D^-$ & $-\frac{1}{3\sqrt{3}f^2} \left( \frac{1}{\sqrt{3}} \gamma (2s-t-u) + m_D^2 \right)$ \\
$\eta \eta_c \to D^0 \bar{D}^0$ & $-\frac{1}{3\sqrt{3}f^2} \left( \frac{1}{\sqrt{3}} \gamma (2s-t-u) + m_D^2 \right)$ \\
$\eta \eta_c \to D_s^+ D_s^-$ & $-\frac{1}{3\sqrt{3}f^2} \left( \frac{1}{\sqrt{3}} \gamma (t+u-2s) - m_D^2 - m_K^2 + m_{\pi}^2\right)$ \\
$\eta \eta_c \to \eta \eta^\prime$ & $0$ \\
$\eta \eta_c \to \eta^\prime \eta^\prime$ & $0$ \\
$\eta \eta_c \to \eta^\prime \eta_c$ & $0$ \\
 & \\
$D^+ D^- \to D^+ D^-$ & $-\frac{1}{3f^2} \left( \psi_3 (s+t-2u) + 2m_D^2 \right)$ \\
$D^+ D^- \to D^0 \bar{D}^0$ & $-\frac{1}{6f^2} \left( \psi_5 (t-u) + (s-u) + 2m_D^2 \right)$ \\
$D^+ D^- \to D_s^+ D_s^-$ & $-\frac{1}{6f^2} \left( \psi_5 (t-u) + (s-u) + 2m_D^2 + m_K^2 - m_{\pi}^2\right)$ \\
$D^+ D^- \to \eta \eta^\prime$ & $-\frac{\sqrt{2}}{36f^2} \left( \gamma(2s-t-u) + 2m_D^2 + 2m_{\pi}^2\right)$ \\
$D^+ D^- \to \eta^\prime \eta^\prime$ & $-\frac{1}{36f^2} \left( \gamma(2s-t-u) + 2m_D^2 + 2m_{\pi}^2\right)$ \\
$D^+ D^- \to \eta^\prime \eta_c$ & $-\frac{1}{3\sqrt{6}f^2} \left( \frac{1}{\sqrt{3}} \gamma (2s-t-u) + m_D^2 \right)$ \\
 & \\
$D^0 \bar{D}^0 \to D^0 \bar{D}^0$ & $-\frac{1}{3f^2} \left( \psi_3 (s+t-2u) + 2m_D^2 \right)$ \\
$D^0 \bar{D}^0 \to D_s^+ D_s^-$ & $-\frac{1}{6f^2} \left( \psi_5 (t-u) + (s-u) + 2m_D^2 + m_K^2 - m_{\pi}^2\right)$ \\
$D^0 \bar{D}^0 \to \eta \eta^\prime$ & $-\frac{\sqrt{2}}{36f^2} \left( \gamma(2s-t-u) + 2m_D^2 + 2m_{\pi}^2\right)$ \\
$D^0 \bar{D}^0 \to \eta^\prime \eta^\prime$ & $-\frac{1}{36f^2} \left( \gamma(2s-t-u) + 2m_D^2 + 2m_{\pi}^2\right)$ \\
$D^0 \bar{D}^0 \to \eta^\prime \eta_c$ & $-\frac{1}{3\sqrt{6}f^2} \left( \frac{1}{\sqrt{3}} \gamma (2s-t-u) + m_D^2 \right)$ \\
 & \\
$D_s^+ D_s^- \to D_s^+ D_s^-$ & $-\frac{1}{3f^2} \left( \psi_3 (s+t-2u) + 2m_D^2 + 2m_K^2 - 2m_{\pi}^2 \right)$ \\
$D_s^+ D_s^- \to \eta \eta^\prime$ & $\frac{\sqrt{2}}{18f^2} \left( \gamma(2s-t-u) + 2m_D^2 + 6m_K^2 - 4m_{\pi}^2\right)$ \\
$D_s^+ D_s^- \to \eta^\prime \eta^\prime$ & $-\frac{1}{9f^2} \left( \gamma(2s-t-u) + 2m_D^2 + 6m_K^2 - 4m_{\pi}^2\right)$ \\
$D_s^+ D_s^- \to \eta^\prime \eta_c$ & $-\frac{2}{3\sqrt{6}f^2} \left( \frac{1}{\sqrt{3}} \gamma (2s-t-u) + m_D^2 + m_K^2 - m_{\pi}^2 \right)$ \\
 & \\
$\eta \eta^\prime \to \eta \eta^\prime$ & $-\frac{2}{9f^2} \left( 4m_K^2 - m_{\pi}^2 \right)$ \\
$\eta \eta^\prime \to \eta^\prime \eta^\prime$ & $\frac{\sqrt{2}}{9f^2} \left( 8m_K^2 - 5m_{\pi}^2 \right)$ \\
$\eta \eta^\prime \to \eta^\prime \eta_c$ & $0$ \\
 & \\
$\eta^\prime \eta^\prime \to \eta^\prime \eta^\prime$ & $-\frac{1}{9f^2} \left( 16m_K^2 - 7m_{\pi}^2 \right)$ \\
$\eta^\prime \eta^\prime \to \eta^\prime \eta_c$ & $0$ \\
 & \\
$\eta^\prime \eta_c \to \eta^\prime \eta_c$ & $0$
\end{longtable*}
}}


\end{document}